\documentstyle[12pt]{article}
\newcommand{\sect}[1]{\setcounter{equation}{0}\section{#1}\indent}

\begin{document}
\def\bbox{{\,\lower0.9pt\vbox{\hrule \hbox{\vrule height 0.2 cm
\hskip 0.2 cm \vrule height 0.2 cm}\hrule}\,}}
\def\a{\alpha}
\def\b{\beta}
\def\g{\gamma}
\def\G{\Gamma}
\def\d{\delta}
\def\D{\Delta}
\def\e{\epsilon}
\def\h{\hbar}
\def\ve{\varepsilon}
\def\z{\zeta}
\def\t{\theta}
\def\vt{\vartheta}
\def\r{\rho}
\def\vr{\varrho}
\def\k{\kappa}
\def\l{\lambda}
\def\L{\Lambda}
\def\m{\mu}
\def\n{\nu}
\def\o{\omega}
\def\O{\Omega}
\def\s{\sigma}
\def\vs{\varsigma}
\def\S{\Sigma}
\def\vphi{\varphi}
\def\av#1{\langle#1\rangle}
\def\pa{\partial}
\def\na{\nabla}
\def\hg{\hat g}
\def\un{\underline}
\def\ov{\overline}
\def\cF{{{\cal F}_2}}
\def\Hsl{H \hskip-8pt /}
\def\Fsl{F \hskip-6pt /}
\def\cFsl{\cF \hskip-5pt /}
\def\ksl{k \hskip-6pt /}
\def\pasl{\pa \hskip-6pt /}
\def\tr{{\rm tr}}
\def\tcF{{\tilde{{\cal F}_2}}}
\def\tg{{\tilde g}}
\def\shalf{\frac{1}{2}}
\def\nn{\nonumber \\}
\def\w{\wedge}
\def\be{\begin{equation}\label{}}
\def\ee{\end{equation}}

\def\cmp#1{{\it Comm. Math. Phys.} {\bf #1}}
\def\cqg#1{{\it Class. Quantum Grav.} {\bf #1}}
\def\pl#1{{\it Phys. Lett.} {\bf B#1}}
\def\prl#1{{\it Phys. Rev. Lett.} {\bf #1}}
\def\prd#1{{\it Phys. Rev.} {\bf D#1}}
\def\prr#1{{\it Phys. Rev.} {\bf #1}}
\def\prb#1{{\it Phys. Rev.} {\bf B#1}}
\def\np#1{{\it Nucl. Phys.} {\bf B#1}}
\def\ncim#1{{\it Nuovo Cimento} {\bf #1}}
\def\jmp#1{{\it J. Math. Phys.} {\bf #1}}
\def\aam#1{{\it Adv. Appl. Math.} {\bf #1}}
\def\mpl#1{{\it Mod. Phys. Lett.} {\bf A#1}}
\def\ijmp#1{{\it Int. J. Mod. Phys.} {\bf A#1}}
\def\prep#1{{\it Phys. Rep.} {\bf #1C}}


\begin{titlepage}
\setcounter{page}{0}

\begin{flushright}
COLO-HEP-98/424 \\
hep-th/9905080 \\
May 1999
\end{flushright}

\vspace{5 mm}
\begin{center}
{\large Tachyon Condensation in Rotated Brane Configurations}
\vspace{10 mm}

{\large S. P. de Alwis\footnote{e-mail: 
dealwis@pizero.colorado.edu}}\\
{\em Department of Physics, Box 390,
University of Colorado, Boulder, CO 80309.}\\
\vspace{5 mm}
\end{center}
\vspace{10 mm}

\centerline{{\bf{Abstract}}}
The decay of  rotated brane configurations
 and the corresponding condensation of tachyons is discussed.  In a certain IIB orbifold case a heuristic argument about the mass of the state living on the fixed plane is made. When the rotation angle is $\pi$ this mass agrees with that obtained by Sen.

\end{titlepage}
\newpage
\renewcommand{\thefootnote}{\arabic{footnote}}
\setcounter{footnote}{0}

\setcounter{equation}{0}            
\sect{Introduction}
In the last year there has been  significant progress in understanding certain non-BPS
states in string theory. In particular it has been shown by Sen \cite{sen1}-\cite{sen4}
that there are unstable non-BPS Dp-branes with even (odd) p in type IIB (IIA) branes
and a stable non-BPS D-particle in type I theory. In the case of IIA and IIB too
one could stabilize the non-BPS branes by dividing out by certain $Z_2$ symmetries.
A stable D-particle state in type IIB was explicitly constructed using the 
boundary state formalism in \cite{bg1}. The relation of these
developments to K-theory has been pointed out in \cite{ew} where 
some new insights have been obtained. These  have been
analyzed using the boundary state formalism in \cite{fgls}\footnote{For recent reviews of these developments
see \cite{sen5}.}.

All the above considerations were made in the context of (anti-) parallel brane configurations.
In this paper we wish to consider situations where branes are not parallel (or anti-parallel). We will begin with a review of the evaluation of the partition function for open strings
with  ends on  two D-branes, one of which is rotated with respect
to the other, and the observation that there is a tachyon in the 
system for general angles when the distance between the branes is less than
a certain critical value. We argue that (at least when the brane directions are compact) this tachyon represents an
instability of the system to decay into a BPS brane which is obtained from
the vector sum of the charges of the original branes. Next we construct boundary states corresponding to such  D-branes and verify that they are consistent with open string
closed string duality. 
We consider, closely following Sen's work \cite{sen2},  this system in type $IIB/Z_2$
where the $Z_2$ is generated by a reflection in four coordinates 
times $(-1)^{F_L}$, 
compactify one direction (which is common to both branes) on a circle,
and put a Wilson line on it.
 Upon tachyon condensation, it is argued that we now  have a solitonic state on the orbifold fixed plane whose mass depends on the angle of rotation
between the D-branes, in addition to a brane that 
stretches between the two fixed planes. When the angle is $\pi$ (corresponding
to one of the branes being an anti-brane) the charge and hence mass of the extended brane is zero, and the mass of the soliton on the fixed plane
is equal to
that calculated by Sen. However unlike in this latter case, for a general
angle  we have not been able to establish these results rigorously,
since
it has not been possible to find a conformal field theory argument.
 
The 
calculations are performed in the double Wick rotated  light
cone gauge formalism adopted in \cite{bg2},\cite{sen2}. Strictly
speaking since in this formalism the time direction is transverse
to the brane (all  directions parallel to the brane have positive
metric) the configurations are really (p+1)-instantons. However we 
will assume (as in \cite{bg2},\cite{sen2}) that this is equivalent to a calculation with p-branes
when done with all the ghost complications. We will take the original time coordinate to be 
$x^2$ and the light cone directions after Wick rotation to be $x^0,x^1$.

\sect{The open string partition function for rotated D-branes}

We begin by quoting a calculation given in  \cite{jp} (section 13.4) for the open string partition function, where strings have one end attached to a three-brane and the other to
another three brane, rotated with respect to the first. Actually  the treatment in \cite{jp} is for a four-brane, but in our  framework this is to be interpreted as a four instanton that
is equivalent to a double Wick rotated 3-brane. One of the 
three branes is oriented along the $x^2,x^4,x^6,x^8$ (it should be remembered that $x^2$ is really time)
directions and the other is rotated with respect to this
by an angle $\phi_1$ in the $x^2-x^3$ plane, $\phi_2$ in the $x^4-x^5$ plane etc with $0<\phi_i <\pi$. In our
convention the first rotation would correspond to  a
Lorentz transformation when we Wick rotate back.
The result of the calculation is
\begin{eqnarray}\label{openrot}
\int {dt\over 2t}\tr_{NS}e^{-2\pi tH_o}&=&\int_0^{\infty}{dt\over t}{e^{-{ty^2\over 2\pi\a '^2}}\over (8\pi^2\a 't)^{\shalf}}\prod_{i =1}^4{\t_{00}({i\phi_i t\over\pi},it)\over\t_{11}({i\phi_i t\over\pi},it)} \nn
\int {dt\over 2t}\tr_{NS}(-1)^Fe^{-2\pi tH_o}&=&\int_0^{\infty}{dt\over t}{e^{-{ty^2\over 2\pi\a '^2}}\over (8\pi^2\a 't)^{\shalf}}\prod_{i =1}^4{\t_{01}({i\phi_i t\over\pi},it)\over\t_{11}({i\phi_i t\over\pi},it)} \nn
\int {dt\over 2t}\tr_Re^{-2\pi tH_o}&=&\int_0^{\infty}{dt\over t}{e^{-{ty^2\over 2\pi\a '^2}}\over (8\pi^2\a 't)^{\shalf}}\prod_{i =1}^4{\t_{10}({i\phi_i t\over\pi},it)\over\t_{11}({i\phi_i t\over\pi},it)} \nn
\int {dt\over 2t}\tr_R(-1)^Fe^{-2\pi tH_o}&=&\int_0^{\infty}{dt\over t}{e^{-{ty^2\over 2\pi\a '^2}}\over (8\pi^2\a 't)^{\shalf}}\prod_{i =1}^4{\t_{11}({i\phi_i t\over\pi},it)\over\t_{11}({i\phi_i t\over\pi},it)}. 
\end{eqnarray}
where we have taken the distance between the branes to be $y$.
Let us specialize to the case when three of the angles are
zero. The $\t_{11}$ function in the denominators  blow up when $\phi =0$ and one needs to substitute $i\t_{11}({i\phi t\over\pi},it)\rightarrow L\eta (it)^{-3}(8\pi^2\a 't)^{-\shalf}$ (equation (13.4.24)of \cite{jp})
for the three parallel directions (each of whose size is taken 
to be $L$). Then we get for the potential, 
\begin{equation}\label{pot}
V=L^3\int_0^{\infty}{dt\over t}(8\pi^2\a 't)^{-2}e^{-ty^2\over 2\pi\a'}{\t_{11}
({i\phi t\over 2\pi},it)^4\over \eta (it)^9\t_{11}
({i\phi t\over \pi},it)}.\end{equation}
The NS ground state energy of the system is  $E^0_{NS}=-\shalf+{\phi\over 2\pi}$ but this is projected out by GSO. The first excited state
(which survives the projection) is $\psi_{-\shalf +{\phi\over\pi}}|0>_{NS}$
and has mass  given by 
\begin{equation}\label{}
m^2={y^2\over 4\pi^2\a'^2}-{\phi\over 2\pi\a '}.\end{equation}
This becomes tachyonic at distances less than a critical distance
$y_c\equiv 2\pi\a '\phi$.
If we look at the large $t$  asymptotics of (\ref{pot}) we find
\begin{equation}\label{asym}
V\simeq -L^3\int_0^{\infty}{dt\over t}(8\pi^2\a 't)^{-2}e^{-{ty^2\over 2\pi\a'}+\phi t} 
\end{equation}
which diverges at $y^2<2\pi\a '\phi$ in agreement with (\ref{asym}).

We note that $\phi =\pi$ is equivalent to replacing one of the 
branes by an anti-brane\footnote{This is because at $\phi=\pi$
the RR exchange term flips sign compared to its value at $\phi=0$}. 
In this case the tachyonic mode reflects the annihilation of the
brane anti-brane pair into the vacuum.

The tachyonic mode represents an instability in the system for distances less than $y_c$. In particular  at zero distance one
expects the state to decay into the BPS state corresponding to
the vector sum of the two RR charges of the branes\footnote{It has been argued in \cite{jp} that in the case when all directions are non-compact, the D-branes reconnect and move apart indefinitely. On the other hand in the compact case there is a lower bound to the energy and the  scenario discussed
in this paper should be valid when we consider the system with the plane in which the rotation is made to be a two torus with radii tending to infinity.}. If the two states have RR charges ${\bf Q}_{1,2}$ with unit magnitude then ${\bf Q}_1.{\bf Q}_2=\cos\phi$, and the 
mass of this state is given by\footnote{This can be established by an argument along the lines of the one given
on page 168 of \cite{jp}.},
\begin{equation}\label{tphi}
M_p(\phi )=L^pT_p|{\bf Q}_1+{\bf Q}_2|=\sqrt 2L^pT_p(1+\cos\phi )^{\shalf}=2L^pT_p\cos{\phi\over 2},\end{equation}
where $T_p$ is the p-brane tension and $L~(\rightarrow\infty)$
is the size of the spatial directions in which the brane is wound. 

Note that the above result makes sense only in the compactified
case where (at least) the plane of rotation is on a two torus. The factor $2\cos{\phi\over 2}$ is just the geometrical  
factor that comes from the length of the final brane if we had started with initial branes of unit length. 

It is possible that our conjecture that the two rotated brane configuration 
decays into the above BPS state can be established along the lines of \cite{gns} and \cite{sen3}. The case considered
in \cite{gns} is one where tachyon condensation on the D0-D2 system (or a Dp-D(p-2) system) gives a D(p-2) brane with magnetic flux on it.
Our case is related to this in that a rotated brane is T-dual to 
one with magnetic flux on it. 
Thus we 
believe that the conjecture is very plausible since the above BPS state is the lowest state with the same quantum numbers as the starting configuration. One puzzling aspect of this however
is that the tachyon would be expected to be localized to the
region of intersection of the two branes, but its condensation
must result in changing the energy density over the whole configuration. This issue of course obtains even in tachyon 
condensation in, for instance, the D2-D0 case, where the tachyon would be
localized at the location of the D0-brane. This is a question that needs to be investigated 
further\footnote{I wish to thank A. Sen for commenting on this.}.

\sect{Boundary states}
The boundary state formalism was used to discuss open closed string duality in a series of papers by the authors of \cite{clny} and in \cite{pc}\footnote{This formalism was extended to the discussion
of D-branes in \cite{li}. For references on the boundary state formalism  which are closely related to the
discussion in this section see \cite{bs1},\cite{bs2},\cite{rs}.}.
Here we will construct boundary states corresponding to branes
at angles  and check open closed string duality.
Instead of working with rotated branes we will turn on a ten dimensional gauge field (with constant field strength $F$) and follow
closely the discussion in \cite{clny}.
 In the D-brane context 
the field strength components in the parallel parallel directions
will be  magnetic fields living on the D-brane while the parallel perpendicular 
field strengths will correspond to a rotation of the brane through
the relation $X^{\perp}=2\pi\a 'F^{\perp}_{\parallel}X^{\parallel}$. Henceforth we will write $2\pi\a 'F\rightarrow F$. 
 
The NSNS boundary state in the 
presence of $F$ is given by 
\begin{equation}\label{nsb}
|F,\eta,k>_{NSNS}=N(F)\det(1+F)^{\shalf}\exp\left\{\sum_{n=1}^{\infty}
{1\over n}\a_{-n}^{\mu} T(F)_{\mu\nu}\tilde\a_{-n}^{\nu}+i\eta\sum_{r={\cal Z}^+-\shalf}\psi_{-r}^{\mu}T_{\mu\nu}\tilde\psi^{\nu}_{-r}\right\}|\eta,k >_{NSNS}^0
\end{equation}
 and the RR sector one by,
\begin{equation}\label{rrb}
|F,\eta,k >_{RR}=4iN(F)\exp\left\{\sum_{n=1}^{\infty}{1\over n}\a_{-n}^{\mu} T(F)_{\mu\nu}\tilde\a_{-n}^{\nu}+i\eta\sum_{r={\cal Z}^+}\psi_{-r}^{\mu}T_{\mu\nu}\tilde\psi^{\nu}_{-r}\right\}|F,\eta,k >_{RR}^0,
\end{equation}
where 
\begin{equation}\label{}
|F,\eta,k >_{RR}^0=\tr\exp\left (-\shalf\psi^{\nu}_{\mp}F_{\mu\nu}\psi^{\mu}_{\pm}\right )|\eta,k >_{RR}^{0},
\end{equation}
and $T(F)={1-F\over 1+F}T_0$ with $T_0$ a diagonal matrix with
$-1$ in the N directions and +1 in the D directions. 
$\eta =\pm$ and  
$\psi_{\pm}={1\over\sqrt 2}(\psi_0\pm i\tilde\psi_0)$
with $\psi_{\eta}^{\parallel}|\eta ,k>_{RR}^0=\psi_{-\eta}^{\perp}|\eta ,k>_{RR}^0=0$.
Note that the Born-Infeld pre-factor in the NSNS state is derived
in \cite{clny} and the relative normalization of the NSNS and
the RR states is a consequence of 
supersymmetry as shown there. The above are obtained from those given
in \cite{clny} by T-duality. The 
absolute normalization will be fixed later by comparison with the open string 
calculation \cite{jp} reviewed in the previous section.
The GSO projected boundary states are then
\begin{equation}\label{}
|U,F>_{NSNS}={1\over\sqrt 2}(|F.+ >_{NSNS}+|F.- >_{NSNS}),
\end{equation}
and 
\begin{equation}\label{Ustate}
|U,F>_{RR}={1\over\sqrt 2}(|F.+ >_{RR}+|F.- >_{RR}).
\end{equation}
The D-p-brane state is then given by
\begin{equation}\label{Dstate}
|D,F>={1\over\sqrt 2}(|U,F>_{NSNS}+|U,F >_{RR}).
\end{equation}

Choosing
a basis where the $F$ field is skew diagonal we may put 
\begin{equation}\label{}
F_{\mu\nu}=\oplus \pmatrix{0&f_i \cr -f_i&0 \cr},~f_i=\tan{\phi_i }
\end{equation}
We take as in the previous section the  coordinates  ${\bf x_{\parallel}}=(x^2,x^4,
x^6,x^8)$ as the directions along one of the branes and ${\bf x_{\perp}}=(x^3,x^5,x^7,x^9)$
as the D directions (with $x^0, x^1$ as light cone directions).

 Defining the projection matrices 
$P_{\pm}(F)=\shalf (1\pm T(F))$, the momentum eigenstate may be written as $|k>=|P_+(F)k_R>$
in terms of the right moving momentum eigenvalue $k_R$. The states
are normalized as $<k|k'>=\d (k'-k)$, and the transverse position 
state is then given by,
\begin{equation}\label{}
|x>=\int dk \d(P_-(F)k)e^{-i(P_+(F)k)^T.x}|k>
\end{equation}
In evaluating the matrix element of $e^{-sH_c}$ between D-brane
states we therefore get integrals of the form, (say in the 2-3 plane)
\begin{equation}\label{}
\int dk_2'dk_3'\int dk_2dk_3\d (k_2')\d [\shalf (1+\cos{2\phi} )k_2+\shalf \sin{2\phi} k_3]
e^{-ix_3k_3'-\shalf\a 'sk_3'^2}\d^2 (k-k')={2\over \sin{2\phi}}
\end{equation}

When $\phi =0,\pi$, we have putting $\d(k'-k)={L\over 2\pi}\d (k_{\perp}'-k_{\perp})$ in the integral a factor,
\begin{equation}\label{}
{L\over 2\pi}\sqrt{2\pi\over \a 's}e^{-{x_{\perp}^2\over 2\a 's}}.
\end{equation}
There is a similar factor associated with the light cone direction.

Putting all this together we have the following results.
Define
\begin{eqnarray}\label{}
\int_s&\equiv&N(0)N(F)\prod_{i=1}^4 {2\over\sin{\phi_i}}2^4\int_0^{\infty} ds\sqrt{2\pi\over\a 's}e^{-y^2\over 2\a 's} \nn
\int_t&\equiv&N(0)N(F)\prod_{i=1}^4 {2\over\sin{\phi_i}}2^4\int_0^{\infty}{\pi dt\over t}\sqrt{2\over\a 't}e^{-y^2t\over 2\a '\pi}. 
\end{eqnarray}
where $y$ is, as in the previous section, the distance between the branes.
Then we have
\begin{eqnarray}\label{stamps1}
\int_0^{\infty} ds<0,\pm |e^{-sH_c}|F,\pm >_{NSNS}&=&\int_s\prod_{i=1}^4\tan{\phi_i}{\t_{00}({\phi_i\over \pi},{is\over\pi})\over\t_{11}({\phi_i\over \pi},{is\over\pi})} \nn
=\int_t\prod_{i=1}^4\tan{\phi_i}{\t_{00}({-it\phi_i\over \pi},it)\over\t_{11}({-it\phi_i\over \pi},it)}&=&\int_0^{\infty}{dt\over 2t}\tr_{NS}e^{-2\pi tH_o}.
\end{eqnarray}
\begin{eqnarray}\label{stamps2}
\int_0^{\infty} ds<0,\pm |e^{-sH_c}|F,\mp >_{NSNS}&=&\int_s\prod_{i=1}^4\tan{\phi_i}{\t_{01}({\phi_i\over \pi},{is\over\pi})\over\t_{11}({\phi_i\over \pi},{is\over\pi})} \nn
=\int_t\prod_{i=1}^4\tan{\phi_i}{\t_{10}({-it\phi_i\over \pi},it)\over\t_{11}({-it\phi_i\over \pi},it)}&=&\int_0^{\infty}{dt\over 2t}\tr_Re^{-2\pi tH_o}.
\end{eqnarray}
\begin{eqnarray}\label{stamps3}
\int_0^{\infty} ds<0,\pm |e^{-sH_c}|F,\pm >_{RR}&=&-\int_s\prod_{i=1}^4\tan{\phi_i}{\t_{10}({\phi_i\over \pi},{is\over\pi})\over\t_{11}({\phi_i\over \pi},{is\over\pi})} \nn
=-\int_t\int_t\prod_{i=1}^4\tan{\phi_i}{\t_{01}({-it\phi_i\over \pi},it)\over\t_{11}({-it\phi_i\over \pi},it)}&=&\int_0^{\infty}{dt\over 2t}\tr_{NS}(-1)^Fe^{-2\pi tH_o}.
\end{eqnarray}
\begin{eqnarray}\label{stamps4}
\int_0^{\infty} ds<0,\pm |e^{-sH_c}|F,\mp >_{RR}&=&-\int_s\prod_{i=1}^4\tan{\phi_i}{\t_{11}({\phi_i\over \pi},{is\over\pi})\over\t_{11}({\phi_i\over \pi},{is\over\pi})} \nn
=-\int_t\prod_{i=1}^4\tan{\phi_i}{\t_{11}({-it\phi_i\over \pi},it)\over\t_{11}({-it\phi_i\over\pi},it)}&=&-\int_0^{\infty}{dt\over t}\tr_R(-1^F)e^{-2\pi tH_o}.
\end{eqnarray}
The second equality in each of the above equations  is the result
of the transformation $s\leftarrow t={\pi\over s}$. The subscripts
on the trace refers to the states (NS or R) over which the trace is 
taken.
The last equality follows from  comparison with the calculation
of the previous section (the open string loop calculation) and 
is valid if we choose the normalization constant to be
$N(F)={1\over 8\sqrt 2\pi}\prod_i\cos^2{\phi_i}$. With this  we have (using (\ref{Dstate}) and (\ref{Ustate}))the 
result
\begin{equation}\label{}
\int_0^{\infty} ds<D,0 |e^{-sH_c}|D,F >=\int_0^{\infty}{dt\over t}\tr_{NS-R}\shalf (1+(-1)^F)e^{-2\pi tH_o}.
\end{equation}
This checks the open string closed string duality for rotated branes.
In particular when three of the phis are zero and one is equal to
$\pi$  we get the result for brane anti-brane interaction given in section 2.1 of \cite{sen2}.

\sect{Rotated D-branes on a $Z_2$ orbifold}

Let us, following Sen \cite{sen2}, consider an orbifold (of IIB string theory) generated by 
\be
g=(-1)^{F_L}I_4,~~I_4:x^{6,7,8,9}\rightarrow -x^{6,7,8,9}\ee
 where $(-1)^F_L$ acts as -1 on the left moving Ramond ground state
and is +1 on all other ground states. 
 For concreteness we again consider a 3-brane but this time we take
\be 
{\bf x_{\parallel}}=\{x^2,x^3,x^4,x^8 \},~~{\bf x_{\perp}}=\{x^5,x^6,x^7,x^9 \}.\ee

The light cone directions are as before $\{x^0,x^1\}$ and we are 
working again in the double Wick rotated formalism where
$x^2$ is the Euclideanized time. We take one brane to be stretched along
the above $x_{\parallel}$ directions and the other to be parallel to it in the 2,3 and 8 directions but rotated 
with respect to it in the $4-5$ plane by an angle $\phi$.
As in \cite{sen2} one can check that 
\be g|U,F>_{NSNS}=|U,F>_{NSNS}, ~~g|U,F>_{RR}=|U,F>_{RR}.\ee
In addition to these untwisted states there are twisted states that 
can be represented as
\begin{eqnarray}\label{}
|F,\eta,k>_{NSNS;T}&=&N(F)\det(1+F)^{\shalf}\nn
& &\exp\left\{\sum_{n}
{1\over n}\a_{-n}^{\mu} T(F)_{\mu\nu}\tilde\a_{-n}^{\nu}+i\eta\sum_{r}\psi_{-r}^{\mu}T_{\mu\nu}(F)\tilde\psi^{\nu}_{-r}\right\}|\eta ,k >_{NSNS;T}^0,\nn 
\end{eqnarray}
\begin{equation}\label{rrb}
|F,\eta,k >_{RR;T}=4iN(F)\exp\left\{\sum_{n}{1\over n}\a_{-n}^{\mu} T(F)_{\mu\nu}\tilde\a_{-n}^{\nu}+i\eta\sum_{r}\psi_{-r}^{\mu}T_{\mu\nu}\tilde\psi^{\nu}_{-r}\right\}|F,\eta ,k >_{RR;T}^0.\end{equation}
In the above sums we take
 \begin{eqnarray}\label{}
NSNS:~~~n&\e & {\cal Z_+}; \mu =2,3,4,5;~~~r\e{\cal Z_+}-\shalf \nn
&\e&{\cal Z_+}-{\shalf};\mu =6,7,8,9;~~r\e{\cal Z_+}, 
\end{eqnarray}
\begin{eqnarray}\label{}
RR:~~~n,r&\e & {\cal Z_+}; \mu =2,3,4,5; \nn
&\e&{\cal Z_+}-{\shalf};\mu =6,7,8,9. 
\end{eqnarray}
Because of the orbifold condition the momentum integration is 
just over the perpendicular directions that are not transverse to
the orbifold fixed plane. So we define
\begin{eqnarray}\label{}
|\eta>_{NSNS;T}&=&2\tilde N\int dk^0dk^1 |k,\eta > \nn
|\eta>_{RR;T}&=&2\tilde N\int dk^0dk^1 |k,\eta >. 
\end{eqnarray}
As in \cite{sen2} the GSO invariant twisted states are then given by 
\begin{eqnarray}\label{Tstate}
|T>_{NSNS}&=&{1\over\sqrt 2}(|+>_{NSNS;T})+|->_{NSNS;T}) \nn
|T>_{RR}&=&{1\over\sqrt 2}(|+>_{RR;T})+|->_{RR;T}). 
\end{eqnarray}
The untwisted matrix elements are given by the same formulae as before (\ref{stamps1}-\ref{stamps4}) with the replacements
appropriate to the case when three of the angles are zero (see remark after equation (2.1)). For the twisted ones we now get,
\begin{eqnarray}\label{}
\int_0^{\infty}ds <0,\pm |e^{-sH_c}|F,\pm>_{NSNS;T}&=&\tilde N(0)\tilde N(F){2\pi\over\a '}\int{dt\over t}{\sin\phi\over \cos\phi}{\t_{00}({-it\phi\over\pi},it)\t_{00}(0,it)\t_{01}(0,it)^2\over\eta (it)^3
\t_{11}({-it\phi\over\pi},it)\t_{10}(0,it)^2}\nn &=&\int_0^{\infty}
{dt\over 2t}\tr_{NS,\phi}e^{-2\pi tH_0}g,\nn
\int_0^{\infty}ds <0,\mp |e^{-sH_c}|F,\pm>_{NSNS;T}
 &=&0=-\int_0^{\infty}{dt\over 2t}\tr_{R,\phi}e^{-2\pi tH_0}g,\nn
\int_0^{\infty}ds <0,\pm |e^{-sH_c}|F,\pm>_{RR;T}&=&\tilde N(0)\tilde N(F){2\pi\over\a '}\int{dt\over t}{\sin\phi\over \cos\phi}{\t_{01}({-it\phi\over\pi},it)\t_{01}(0,it)\t_{00}(0,it)^2\over\eta (it)^3
\t_{11}({-it\phi\over\pi},it)\t_{10}(0,it)^2}\nn
&=&\int_0^{\infty}{dt\over 2t}\tr_{NS,\phi}e^{-2\pi tH_0}(-1)^F.g,\nn
\int_0^{\infty}ds <0,\mp |e^{-sH_c}|F,\pm>_{RR;T}&=&0=-\int_0^{\infty}{dt\over 2t}\tr_{R,\phi}e^{-2\pi tH_0}(-1)^F.g,
\end{eqnarray}
We have performed the transformation $s={\pi\over t}$ in the
expressions above and then compared with the corresponding open string
calculation. As in the previous discussion these equations will also
thus fix the overall normalization constant $\tilde N(F)$.

The complete D-brane state is then defined (after changing 
the notation slightly  by replacing $F$ in the state by the angle of rotation $\phi$)
\be |D;\phi ,\pm > = \shalf[|U,\phi >_{NSNS}+|U,\phi >_{RR}\pm (|T,\phi >_{NSNS}+|T,\phi >_{RR})\ee
where the states on the right hand side are defined in (\ref{Ustate})
and (\ref{Tstate}). It should be noted that the anti-brane state is
given as $|\overline D;0,\pm >=|D;\pi,\pm>$. Using the above results
we then have
\begin{eqnarray}\label{}
\int_0^{\infty}ds<D;0\pm|e^{-sH_c}|D;\phi ,\pm>&=&\int_0^{\infty}{dt\over 2t}\tr_{NS-R,\phi}e^{-2\pi tH_0}\shalf (1+(-1)^F).\shalf (1+g)~. \nn
\int_0^{\infty}ds<D;0\pm |e^{-sH_c}|D;\phi ,\mp >&=&\int_0^{\infty}{dt\over 2t}\tr_{NS-R,\phi}e^{-2\pi tH_0}\shalf (1+(-1)^F).\shalf (1-g)~.\nn 
\end{eqnarray}
Note that the effect of replacing one of the branes by an anti-brane
i.e. $\phi =\pi$ on the right hand side of this is equivalent to putting $\phi =0$ but changing the sign in front of $(-1)^F$.

Now still following Sen \cite{sen2} we compactify the eight direction
on a circle of radius R and put a Wilson line on one of the D-branes.
The wave function then acquires a phase  $e^{i\chi}$ (say) as one goes
round the circle i.e. $x^8\rightarrow x^8+2\pi R$, and $g$ invariance 
restricts the values of $\chi$ to $0,\pi$. Now in the compactified 
space the orbifold action gives two fixed planes at $x^8=0,\pi R$. So it
is convenient to start with a circle of radius $2R$ and introduce two
groups \cite{sen2}, 
\begin{eqnarray}\label{}
g_1&:&~~x^6\rightarrow -x^6,~~x^7\rightarrow -x^7,~~x^8\rightarrow -x^8
~~x^9\rightarrow -x^9,~~and~(-1)^{F_L}  \nn
g_2&:&~~x^6\rightarrow -x^6,~~x^7\rightarrow -x^7,~~x^8\rightarrow -x^8
+2\pi R~~x^9\rightarrow -x^9,~~and~(-1)^{F_L}. 
\end{eqnarray}
Then we also have
\be g_2g_1: x^8\rightarrow x^8+2\pi R.\ee
so that modding out by this symmetry gives us back a circle of radius
$R$. Now we may construct boundary states for this configuration by
adding to the untwisted states the two twisted sectors at the two
orbifold planes.
\begin{eqnarray}\label{state}
|D;\chi,\phi,\e >&=&\shalf [|U;\chi ,\phi >_{NSNS}+|U;\chi,\phi >_{RR} +\e{1\over\sqrt 2}(|T_1 ,\phi >_{NSNS}+|T_1,\phi >_{RR})\nn
&+&e^{i\chi}\e{1\over\sqrt 2}(|T_2 ,\phi >_{NSNS}+|T_2,\phi >_{RR})], 
\end{eqnarray}
with $\e=\pm$. To compare with the corresponding boundary state that Sen constructed \cite{sen2} we note  that for $\phi =0,\pi,$ $|U;\chi,\phi >_{RR}=\e_1(\phi )|U;\chi,0 >_{RR}$ and $|T_{1,2},\phi >_{NSNS}=|T_{1,2},0 >_{NSNS}$, $|T_1,\pi >_{RR}=\e_1|T_1,0 >_{RR}$ where $\e_{1,2}(\phi )=\pm 1$ for $\phi =0,\pi$ respectively   and $\e =\e_2$.

Now we may compute the interaction between two such states
as before to get,
\be \int_0^{\infty}ds<D;0,0,0|e^{-sH_c}|D;\chi,\phi,\e>=\int{dt\over 2t}
\tr_{NS-R,\phi}e^{-2\pi tH_0}{1\over 8}(1+(-1)^F).(1+\e g_1).(1+e^{i\chi}\e g_2).\ee

\sect{Tachyon condensation}
Let us label the D-branes by the coefficients $q_1=\e,~q_2=\e e^{i\chi}$ of the twisted
sector states $|T1,2>$ at each orbifold plane. There are 
four possibilities for $(q_1,q_2)$; a)(+,+), b)(-,-), c)(+,-), d)(-,+).  If we
take a) and c) together then the masses (per unit
volume in the fixed plane directions) in the ground state, after tachyon condensation, of the states
at the two ends $0, \pi R$ of the compact direction may be written as
 $m_{++}$ at one end and $m_{+-}$ at the other. 
This is a situation with a Wilson line ($\chi =\pi$) at one end.
The tachyon field is then anti-periodic in the eighth direction and the corresponding momentum is quantized in half integral units $k_8={n+\shalf\over R}$.
The lowest mode then has mass 
\begin{equation}\label{}
m^2(\phi)={1\over 4R^2}-{\phi\over 2\pi\a '}
\end{equation}
This becomes marginal at 
the critical radius 
\begin{equation}\label{}
R_c=\sqrt{\pi\a '\over 2\phi}
\end{equation}
As in \cite{sen2} at this radius the mass (per unit volume in the fixed plane directions) of the configuration may be taken to be
that of the two branes (taken now to be p-branes), 
\begin{equation}\label{mc}
M_c(\phi )=2\pi T_pR_c =2\pi{1\over g(2\pi )^p(\sqrt\a ')^{p+1}}\sqrt{\pi\a '\over 2\phi}={1\over g(2\pi )^{(p-1)}(\sqrt\a ')^{p}}\sqrt{\pi\over 2\phi}=\sqrt{\pi\over 2\phi}T_{p-1}.
\end{equation} 
 The above depends on the assumption that at the critical radius $R_c$ (above which a tachyonic mode appears) the original configuration becomes degenerate with the final one. This is 
tantamount to the assumption that the tachyon is exactly marginal
at the critical radius. Unfortunately unlike in the brane anti-brane case where there is a  conformal field theory proof \cite{sen4}, for 
general angles we are unable to do better than the above  heuristic argument.

Now in section 2 we argued that the unstable configuration of 
two non-parallel D branes in flat space decays into a BPS state with mass given by (\ref{tphi}). Let us write $m_s=M_p(\phi)\pi R/L^p$. In the present case the tachyon has kinks at the
orbifold fixed planes $x^8=0,\pi R$ (see \cite{sen2}) and 
correspondingly we expect solitonic states localized on the fixed planes. Then  for 
radius R the mass per unit (p-1) volume (transverse to the compactified
direction) of the ground state configuration is conjectured
 to be given by,

\begin{eqnarray}\label{}
m_{++}+m_{+-}+m_s&=&2\pi R_cT_{p-1}+\pi (R-R_c)T_p\sqrt 2(1+\cos\phi )^{\shalf}\nn
&=&T_{p-1}\sqrt{\pi\over 2\phi}(1-\cos{\phi\over 2})+\pi RT_p\sqrt2 (1+\cos\phi )^{\shalf}.
\end{eqnarray}
This expression is constructed so that it gives the formula obtained
above (\ref{mc}) at the  critical radius $R_c$. At $\phi =0$ it reduces to
 twice the mass of the two p-branes and at $\phi =\pi$, the  brane
anti-brane case, one gets $T_{p-1}/\sqrt{2}$ the mass derived by 
Sen \cite{sen2},\cite{sen4}. The second term is the mass of the 
BPS state stretched between the two orbifold planes and so we 
expect the first term to be the total mass of the states localized on the orbifold planes. i.e.
\be\label{ac} m_{++}+m_{+-}=T_{p-1}\sqrt{\pi\over 2\phi}(1-\cos{\phi\over 2})\ee

Now let us consider the combination of the branes a) and b).
In this case there is no Wilson line but only states odd under
$g_1$ and $g_2$ are projected in. The $k_8=0$ mode is projected out and the lowest open string mode
has $k_8=\pm{1\over R}$ and its mass is given by,
\be m^2=k_8^2-{\phi\over 2\pi\a '}\ee 
giving a critical radius of
\be R'_c=\sqrt{2\pi\a '\over\phi}=2R_c.\ee
In this situation we have a state of mass $m_{+-}$ at each end.
Reasoning as before this leads to the equation 
\be\label{ab} m_{+-}= T_{p-1}\sqrt{\pi\over 2\phi}(1-\cos{\phi\over 2})\ee
Using (\ref{ac}) this gives
\be m_{++}=0\ee
Note that for $\phi =0$, $m_{+-}=0$ and for $\phi =\pi$ (the brane-anti-brane case)
we get the value obtained by Sen \cite{sen2}. 

\sect{Comments}
The twisted state RR charge of the final configuration must be
 equal to
the twisted sector charge of the initial configuration at each end.
The final brane stretched between the two orbifold planes also has twisted sector charge at each end and if this does not match the initial configuration charges at either  end then there must be a 
solitonic state at that end.  Now at the
$x^8=0$ end one may reason as follows. The twisted sector charge 
vector at this end is proportional to the untwisted sector charge,
and given the untwisted charges $\bf Q_1, Q_2$ of the two
initial branes the total twisted sector charge of the initial configuration would be proportional to  $\bf Q_1+Q_2$ (see equation (\ref{state}) and the discussion at the begining of section 5). But the 
untwisted sector charge of the final brane stretched between
the two fixed planes is $\bf Q_1 +Q_2$
and hence its twisted sector charge at the $x^8=0$ end is
equal to that of the initial configuration. Hence there is
no residual twisted sector charge at this end and one would 
expect the mass of a ++ state at this end to be zero. This is
consistent with the above result $m_{++}=0$.

At the $x^8=\pi R$ end however the the total twisted sector RR charge 
of the initial branes is proportional to $\bf Q_1-Q_2$ (we have taken the
Wilson line $e^{i\chi}=1$ for the first brane and -1 for the second). But the 
untwisted sector charge of the final brane is the sum of the two
charges and hence its twisted sector charge is proportional to $\pm\bf
(Q_1+Q_2)$ (here we have also used the sign ambiguity coming from the possibility of assigning either $\e =\pm $ to the end of the final brane). Thus (for either sign) there is a discrepancy in the twisted sector charges
at this end which must therefore indicates the presence of a 
$p-1$ dimensional soliton living at this end carrying charge proportional to $2\bf Q_1$ or $2\bf Q_2$. This accounts for
the state with non-zero mass $m_{+-}$ obtained above.
The fact that this charge is independent of the angle $\phi$ while 
the mass $m_{+-}$ depends on the angle presumably means  that
this $p-1$ dimensional  soliton is bound to the  $p$ dimensional brane
and the angular dependence reflects the binding energy\footnote{This section is largely the result of comments made by an annonymous referee}.

\sect{Acknowledgments} 
I wish to thank J. Polchinski and A. Sen for helpful correspondence and the latter
also for comments on the manuscript.
This work is partially supported by
the Department of Energy contract No. DE-FG02-91-ER-40672.


\end{document}